\documentclass{elsart}
\usepackage{epsfig}

\begin{document}

\begin{frontmatter}
\title{Measurement of the proton spectra from Non-Mesonic Weak Decay of $\mathrm{^5_{\Lambda}He}$, 
$\mathrm{^7_{\Lambda}Li}$ and $\mathrm{^{12}_{\Lambda}C}$}

\centering{The FINUDA Collaboration}

\author[polito,infnto]{M.~Agnello},
\author[peter]{A.~Andronenkov},
\author[victoria]{G.~Beer},
\author[lnf]{L.~Benussi},
\author[lnf]{M.~Bertani},
\author[korea]{H.C.~Bhang},
\author[lnf]{S.~Bianco},
\author[unibs,infnpv]{G.~Bonomi},
\author[unitos,infnto]{E.~Botta},
\author[units,infnts]{M.~Bregant},
\author[unitos,infnto]{T.~Bressani},
\author[unitos,infnto]{S.~Bufalino\thanksref{corresponding}},
\author[unitog,infnto]{L.~Busso},
\author[infnto]{D.~Calvo},
\author[units,infnts]{P.~Camerini},
\author[enea]{M.~Caponero},
\author[uniba,infnba]{B.~Dalena},
\author[unitos,infnto]{F.~De~Mori},
\author[uniba,infnba]{G.~D'Erasmo},
\author[infnba]{D.~Elia},
\author[lnf]{F.~L.~Fabbri},
\author[unitog,infnto]{D.~Faso},
\author[infnto]{A.~Feliciello},
\author[infnto]{A.~Filippi},
\author[uniba,infnba]{E.~M.~Fiore},
\author[infnpv]{A.~Fontana},
\author[tokyo]{H.~Fujioka},
\author[lnf]{P.~Gianotti},
\author[infnts]{N.~Grion},
\author[lnf]{O.~Hartmann},
\author[korea]{B.~Kang},
\author[jinr]{A.~Krasnoperov},
\author[korea]{Y.~Lee},
\author[infnba]{V.~Lenti},
\author[lnf]{V.~Lucherini},
\author[infnba]{V.~Manzari},
\author[unitos,infnto]{S.~Marcello},
\author[tokyo]{T.~Maruta},
\author[teheran]{N.~Mirfakhrai},
\author[unipv,infnpv]{P.~Montagna},
\author[cnr,infnto]{O.~Morra},
\author[kyoto]{T.~Nagae},
\author[tokyo]{D.~Nakajima}, 
\author[riken]{H.~Outa},
\author[lnf]{E.~Pace},
\author[uniba,infnba]{M.~Palomba},
\author[infnba]{A.~Pantaleo},
\author[infnpv]{A.~Panzarasa},
\author[infnba]{V.~Paticchio},
\author[units]{S.~Piano},
\author[lnf]{F.~Pompili},
\author[units,infnts]{R.~Rui},
\author[kek]{M.~Sekimoto}, 
\author[uniba,infnba]{G.~Simonetti},
\author[jinr]{V.~Tereshchenko},
\author[kek]{A.~Toyoda},
\author[infnto]{R.~Wheadon},
\author[unibs,infnpv]{A.~Zenoni}

\thanks[corresponding]{corresponding author. e-mail: bufalino@to.infn.it; fax:
+39.011.6707324.}

\address[polito]{Dip. di Fisica, Politecnico di Torino, Corso Duca degli Abruzzi 24, Torino, Italy}
\address[infnto]{INFN Sez. di Torino, via P. Giuria 1, Torino, Italy}
\address[peter] {Department of Physics, St. Petersburg State University, St. Petersburg, Russia}
\address[victoria]{University of Victoria, Finnerty Rd.,Victoria, Canada}
\address[lnf]{Laboratori Nazionali di Frascati dell'INFN, via E. Fermi 40, Frascati, Italy}
\address[korea]{Dep. of Physics, Seoul National Univ., 151-742 Seoul, South Korea}
\address[unibs]{Dip. di Meccanica, Universit\`a di Brescia, via Valotti 9, Brescia, Italy}
\address[infnpv]{INFN Sez. di Pavia, via Bassi 6, Pavia, Italy}
\address[unitos]{Dip. di Fisica Sperimentale, Universit\`a di Torino, via P. Giuria, 1 Torino, Italy}
\address[units]{Dip. di Fisica Universit\`a di Trieste, via Valerio 2, Trieste, Italy}
\address[infnts]{INFN Sez. di Trieste, via Valerio 2, Trieste, Italy}
\address[unitog]{Dip. di Fisica Generale, Universit\`a di Torino, via P. Giuria 1, Torino, Italy}
\address[enea]{ENEA C.R. Frascati, via E. Fermi 45, Frascati, Italy}
\address[uniba]{Dip. InterAteneo di Fisica, via Amendola 173, Bari, Italy}
\address[infnba]{INFN Sez. di Bari, via Amendola 173, Bari, Italy }
\address[tokyo]{Dep. of Physics, Univ. of Tokyo, Bunkyo, Tokyo 113-0033, Japan}
\address[jinr]{JINR, Dubna, Moscow region, Russia}
\address[teheran]{Dep of Physics, Shahid Behesty Univ., 19834 Teheran, Iran}
\address[unipv]{Dip. di Fisica Teorica e Nucleare, Universita' di Pavia, via Bassi 6, Pavia, Italy}
\address[cnr]{INAF-IFSI Sez. di Torino, C.so Fiume 4, Torino, Italy}
\address[kyoto]{Department of Physics, Sakyo-ku, Kyoto 606-8502, Japan}
\address[riken]{RIKEN, Wako, Saitama 351-0198, Japan}
\address[kek]{High Energy Accelerator Research Organization (KEK), Tsukuba, Ibaraki
305-0801, Japan}



\begin{abstract}
The results of a measurement of the proton spectra following the Non--Mesonic Weak Decay of $\mathrm{^5_{\Lambda}He}$, 
$\mathrm{^7_{\Lambda}Li}$ and $\mathrm{^{12}_{\Lambda}C}$ are presented and discussed. The experiment was performed at the ($e^+$ $e^-$) collider DA$\Phi$NE at Laboratori Nazionale di Frascati of INFN.\\ It is the first measurement for $\mathrm{^7_{\Lambda}Li}$, and for all the spectra the lower limit on the energy of the protons is 15 MeV, never reached before. All the spectra show a similar shape, namely a peak at around 80 MeV  as expected for the free $\Lambda p \rightarrow np$ weak reaction, with a low energy rise that should be due to Final State Interactions and/or two--nucleon induced weak processes. The decay spectrum of $\mathrm{^5_{\Lambda}He}$ is somehow similar to the ones reported by previous measurements and theoretical calculations, but the same doesn't happen for the $\mathrm{^{12}_{\Lambda}C}$ one.
\end{abstract}

\end{frontmatter}

\section{Introduction}
The importance of the Non Mesonic Weak Decay (NMWD) of Hypernuclei was recognized since the early days of  Hypernuclear Physics \cite{ref:Ches}. It is the most spectacular example of Nuclear Medium modification (defined by variation of the intrinsic properties of an elementary particle when embedded in a nucleus), even though such a feature was scarcely emphasized. Furthermore it is the easiest way to get information of the four baryon weak process $\Lambda \cal{N} \rightarrow \cal{N} \cal{N}$.\\
In spite of these very appealing features, the process of NMWD of Hypernuclei has been scarcely studied on the experimental side for a few decades, essentially due to experimental difficulties. Indeed it was not only necessary to identify Hypernuclei in their ground states (using a magnetic spectrometer) but also to detect and measure the energy of the emitted protons and neutrons.\\
The first complete measurement was performed in 1991 by Szymanski et al. \cite{ref:BNL91} at BNL using the ($K^{-}$,$\pi^{-}$) reaction at 800 MeV/c to produce and tag $\mathrm{^5_{\Lambda}He}$ and $\mathrm{^{12}_{\Lambda}C}$ hypernuclei, with limited statistics. Proton spectra emitted from  $\mathrm{^{12}_{\Lambda}C}$ formed through the ($\pi^{+}$,$K^{+}$) reaction with $\pi^{+}$ of 1.05 GeV/c at KEK were measured by Noumi et al. \cite{ref:BNL95}.\\
However, the largest amount of data on NMWD was produced by the SKS Collaboration at KEK. The ($\pi^{+}$,$K^{+}$) reaction was used to produce abundantly Hypernuclei in their ground state, identified by the SKS spectrometer \cite {ref:hashi,ref:kim03,ref:Okal,ref:sato,ref:Kan,ref:kim}.

 The main emphasis on the recent mesurements performed with the SKS spectrometer was on the experimental determination of the ratio of the neutron induced NMWD width $\Gamma_{\it n}$ to the proton induced one $\Gamma_{\it p}$. There has been a long standing ''puzzle'' concerning  the $\Gamma_{\it n}/\Gamma_{\it p}$ ratio. The values reported by the former experiments, even though with large errors, were close to or larger than one, whereas the simple one-pion exchange model with $\Delta$I=1/2 rule foresaw  values one order of magnitude lower. In \cite{ref:Kan} a value of about 0.5 was measured in agreement with recent theoretical values obtained with improved one-meson exchange models \cite{ref:parreno} and direct-quark exchange mechanism \cite{ref:Sasa}.
Quite recent review articles on the subject are due to Outa \cite{ref:Ouvar} for the experimental aspects and to Alberico and Garbarino \cite{ref:alberico} for the theoretical approaches.\\ It must be noticed that, if on one side the most important questions on NMWD were qualitatively solved, discrepancies still exist between theory and experiments. As outlined by Bauer et al. \cite{ref:Bau} the experimental proton spectra are incompatible with those evaluated from theoretical models. For this reason we have done an analysis of the proton spectra from NMWD of $\mathrm{^5_{\Lambda}He}$, $\mathrm{^7_{\Lambda}Li}$ and $\mathrm{^{12}_{\Lambda}C}$ taking advantage of the main features of the FINUDA spectrometer, i.e. the trasparency, the use of thin nuclear targets and the measurement of the proton energies by magnetic analysis. A preliminary analysis of the data for $\mathrm{^{12}_{\Lambda}C}$  was already published \cite{ref:HYP}.

\section{Experimental and analysis techniques}

The experiment was performed with the FINUDA spectrometer installed at one of the two interaction regions of the ($e^+$ $e^-$) collider DA$\Phi$NE at LNF. Details of the detector can be found in \cite{ref:FINUDAcarbonio,ref:FINUDA}.\\
The present analysis was done on events collected in the 2003--2004 run (for a total integrated luminosity provided by the collider of 190 pb$^{-1}$) from three $\mathrm{^{12}C}$ targets and in the 2006-2007 run (for a total integrated luminosity provided by the collider of 960 pb$^{-1}$) from two $\mathrm{^{6}Li}$ targets and two $^{7}$Li targets. The thicknesses of the targets were 4 mm for $\mathrm{^{6}Li}$ (90 \% enriched) and $\mathrm{^{7}Li}$ (natural isotopic composition) targets and 2mm for $\mathrm{^{12}C}$  targets (natural isotopic composition).\\
The best momentum resolution achieved up to now for $\pi^{-}$ from Hypernuclei formation ($p_{\pi^{-}}$$\sim$ 260--280 MeV/c)

\begin {equation}
K^{-}_{stop} + ^{A}Z \rightarrow  ^{A}_{\Lambda}Z + \pi^{-}
\label{eq1} 
\end {equation} 

is 0.5\% FWHM \cite{ref:FINUDAcarbonio}. However this resolution was obtained with top quality tracks that were identified by four points ({\it long tracks}) given by the tracking detectors immersed in the 1.0 T magnetic field provided by the superconducting solenoid of FINUDA. In the analysis of the proton spectra from NMWD we required the presence of two particles emitted in coincidence, a $\pi^{-}$ carrying the information of the hypernuclear state (ground state or excited state containing a hypernucleus of lower mass in the ground state) and a proton coming from the same $K^{-}$ interaction primary vertex.\\
In the preliminary analysis \cite{ref:HYP} we required   {\it long tracks} for the determination of the momenta of both $\pi^{-}$ and proton, but we relaxed the criteria for their quality. As a consequence the momentum resolution worsened by a factor of $\sim$2 for $\pi^{-}$ around 250 MeV/c and protons around 400 MeV/c. On the other hand, the statistics on the inclusive $\pi^{-}$ spectra increased by a factor 6. Furthermore only protons with momenta larger than 200 MeV/c, corresponding to an energy threshold of 25 MeV could be detected. The acceptance function for protons was calculated with simulated tracks and taking into account only geometrical effects.\\
In the present analysis we required for the measurement of the proton momenta even tracks reconstructed by means of three points only ({\it short tracks}) provided by the inner detectors (two arrays of Si microstrip detectors and the two layers of Drift Chambers) \cite{ref:agn}. The threshold on the detection energy of protons was then lowered to 15 MeV.\\
Particular care was devoted to the determination of the acceptance function, in particular for low energy protons, which suffer large energy losses in the thin materials they cross along their path (windows of detectors, targets,...). In \cite{ref:HYP} the acceptance function was calculated for simulated tracks and taking into account only the geometrical effects. The new acceptance function, for each target, was evaluated taking into account, besides the geometrical effects, the efficiency of the FINUDA pattern recognition, the trigger request and the quality cuts. The particle identification (PID)for protons was performed by using the information from the Si microstrip arrays and the Drift Chambers. The measured PID efficiency for protons was 95\%.

\section{Experimental results}

\begin{figure}[htb]
\begin{center} 
\begin{tabular}{cc}
\resizebox {7cm}{!}{\includegraphics{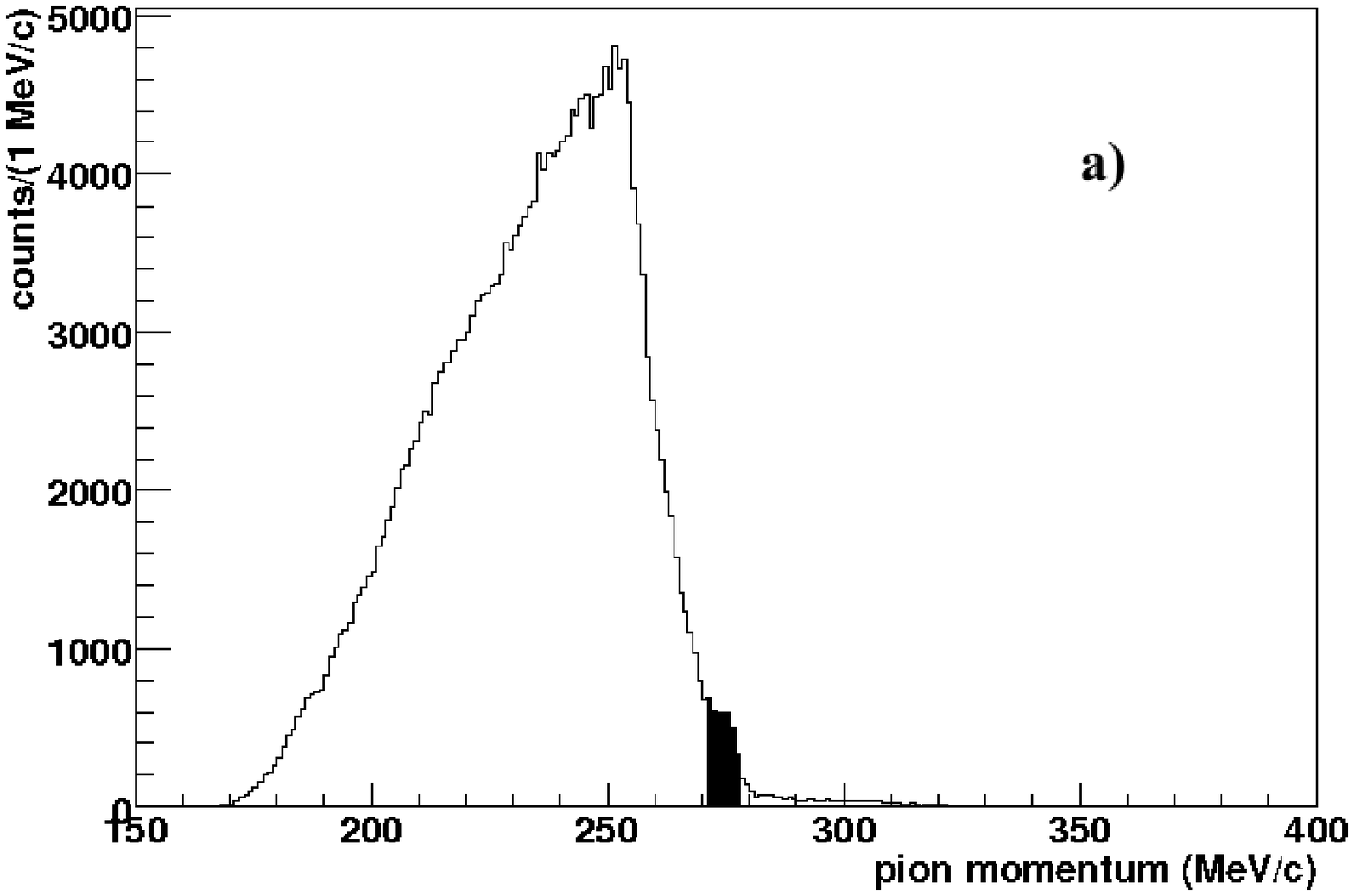}} & \resizebox {7cm}{!}{\includegraphics{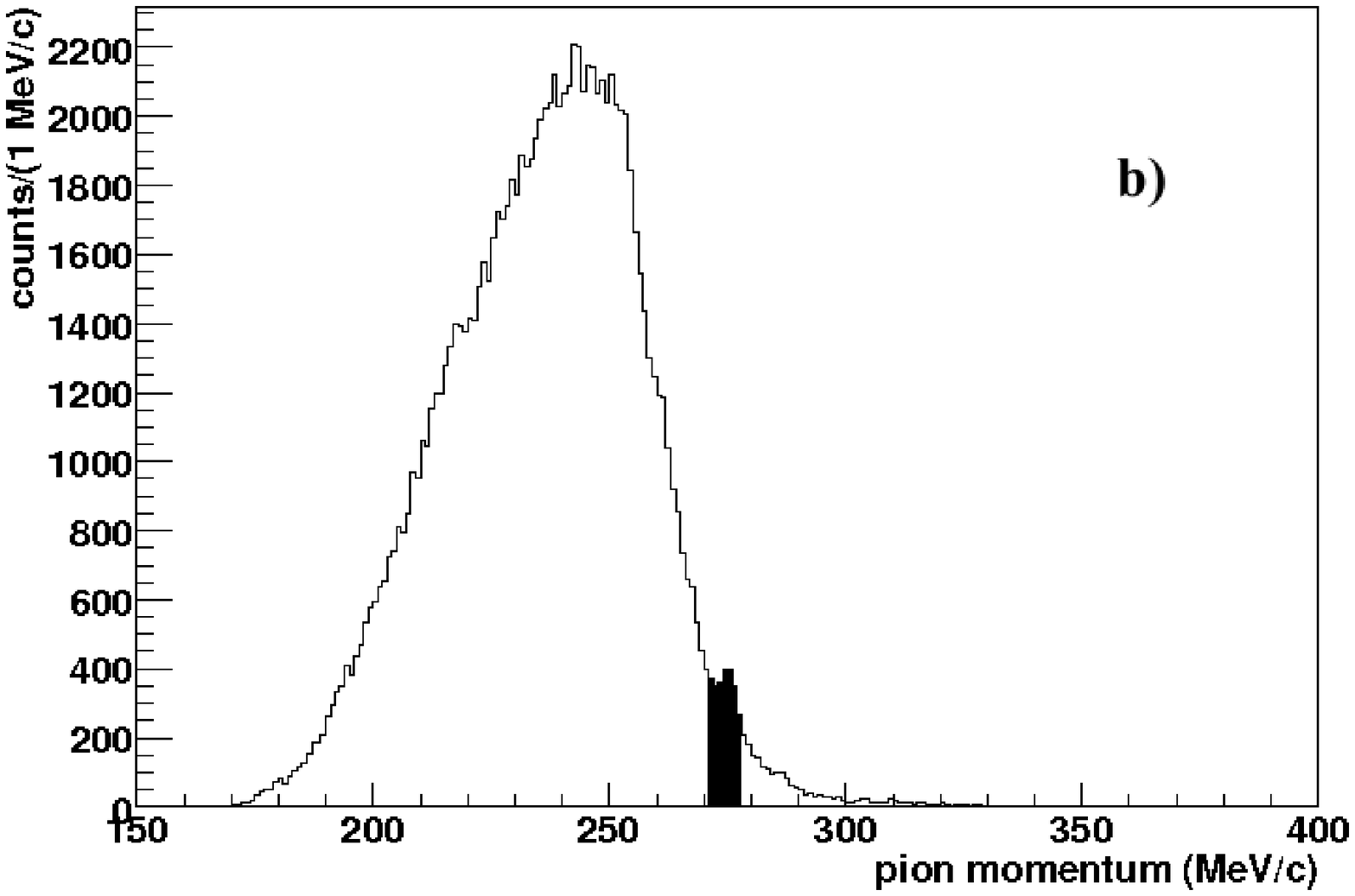}}
\end{tabular}
\resizebox {7cm}{!}{\includegraphics{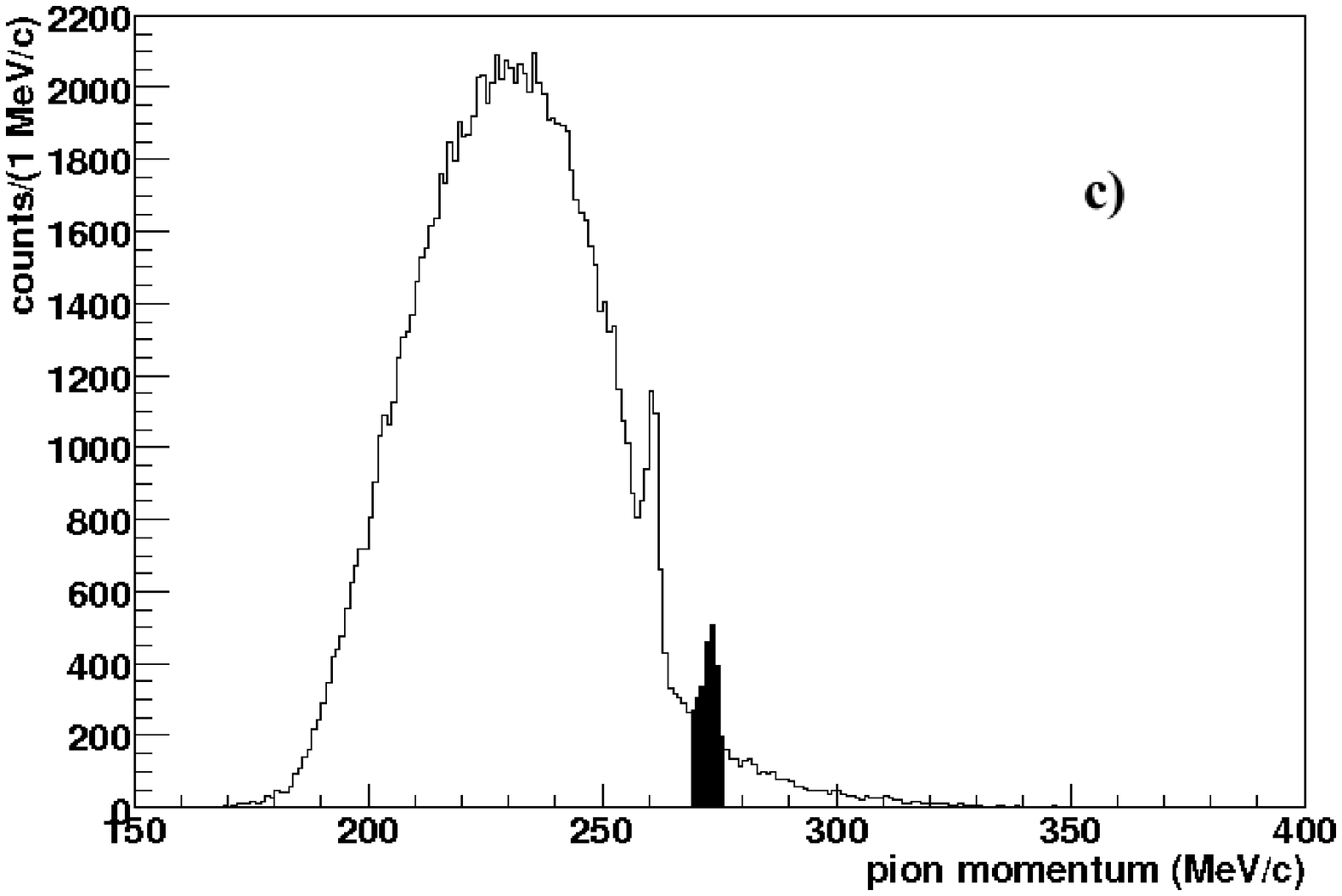}}
\caption{Inclusive momentum spectrum of $\pi^{-}$ emitted from a)$\mathrm{^{6}Li}$ targets, b)$\mathrm{^{7}Li}$ targets and c)$\mathrm{^{12}C}$ targets.}
\label{f:inclusivo}
\end{center}
\end{figure}

Fig.\ref{f:inclusivo} shows the inclusive $\pi^{-}$ spectra measured on the $\mathrm{^{6}Li}$,  $\mathrm{^{7}Li}$ and $\mathrm{^{12}C}$ targets. The black areas indicate the regions in the $\pi^{-}$ spectra (6 MeV/c range) where we expect formation of ($\mathrm{^5_{\Lambda}He_{g.s.}}$+p), $\mathrm{^7_{\Lambda}Li_{g.s.}}$ and $\mathrm{^{12}_{\Lambda}C_{g.s.}}$. A further requirement that improves the signal to background ratios, with respect to the spectrum obtained in \cite{ref:HYP}, is that of a distance less than 3 mm between the primary vertex (interaction of the $K^{-}$) and the secondary vertex of the free $\Sigma^{-}$ hyperon decay process:

\begin {equation}
 \Sigma^{-} \rightarrow  {\it n} +  \pi^{-}
\label{eq2} 
\end {equation} 

 which constitutes the physical background below the peaks due to Hypernuclei formation. We will return on this point later.\\

\begin{figure}[htb]
\begin{center} 
\begin{tabular}{cc}
\resizebox {7cm}{!}{\includegraphics{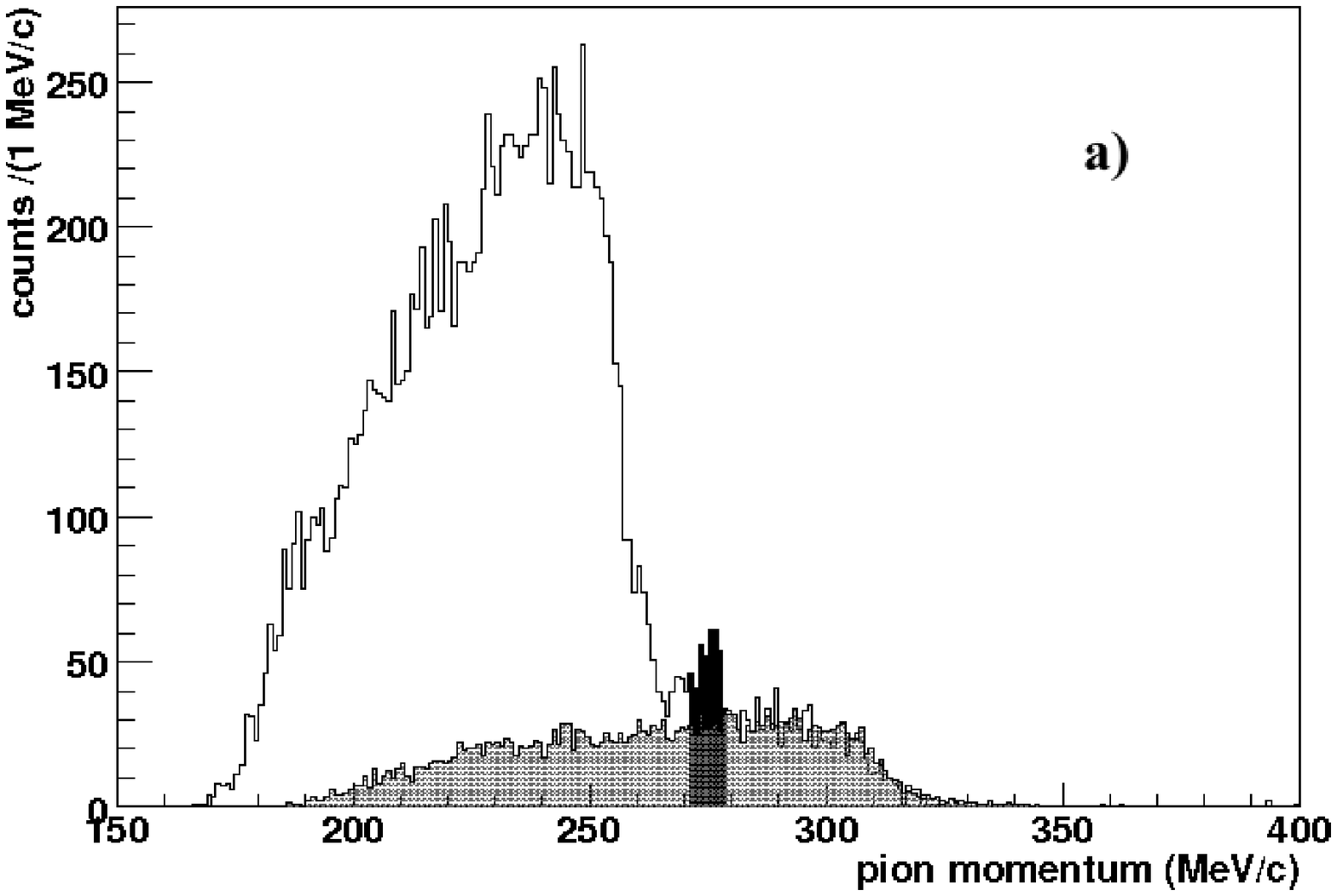}} & \resizebox {7cm}{!}{\includegraphics{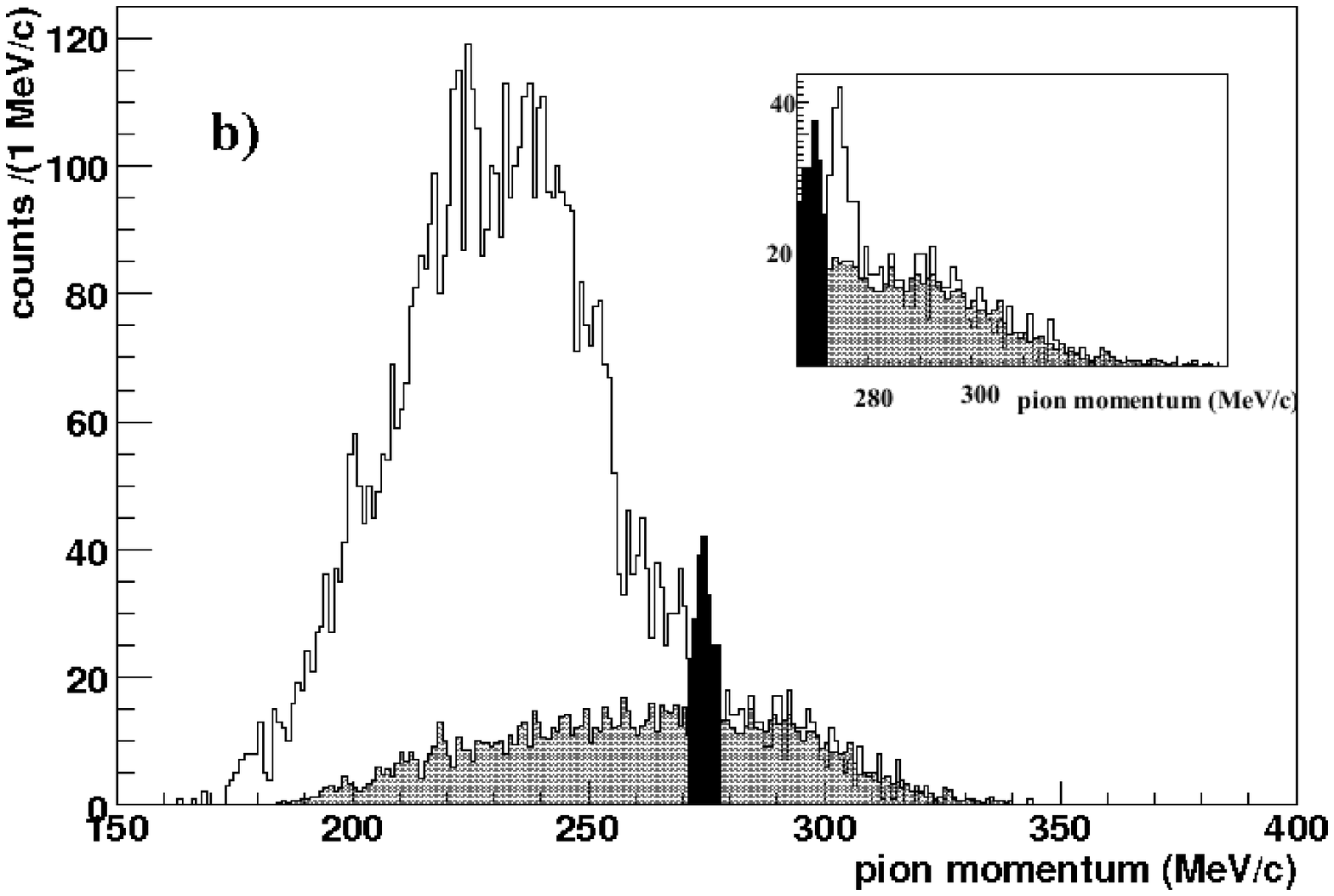}}
\end{tabular}
\resizebox {7cm}{!}{\includegraphics{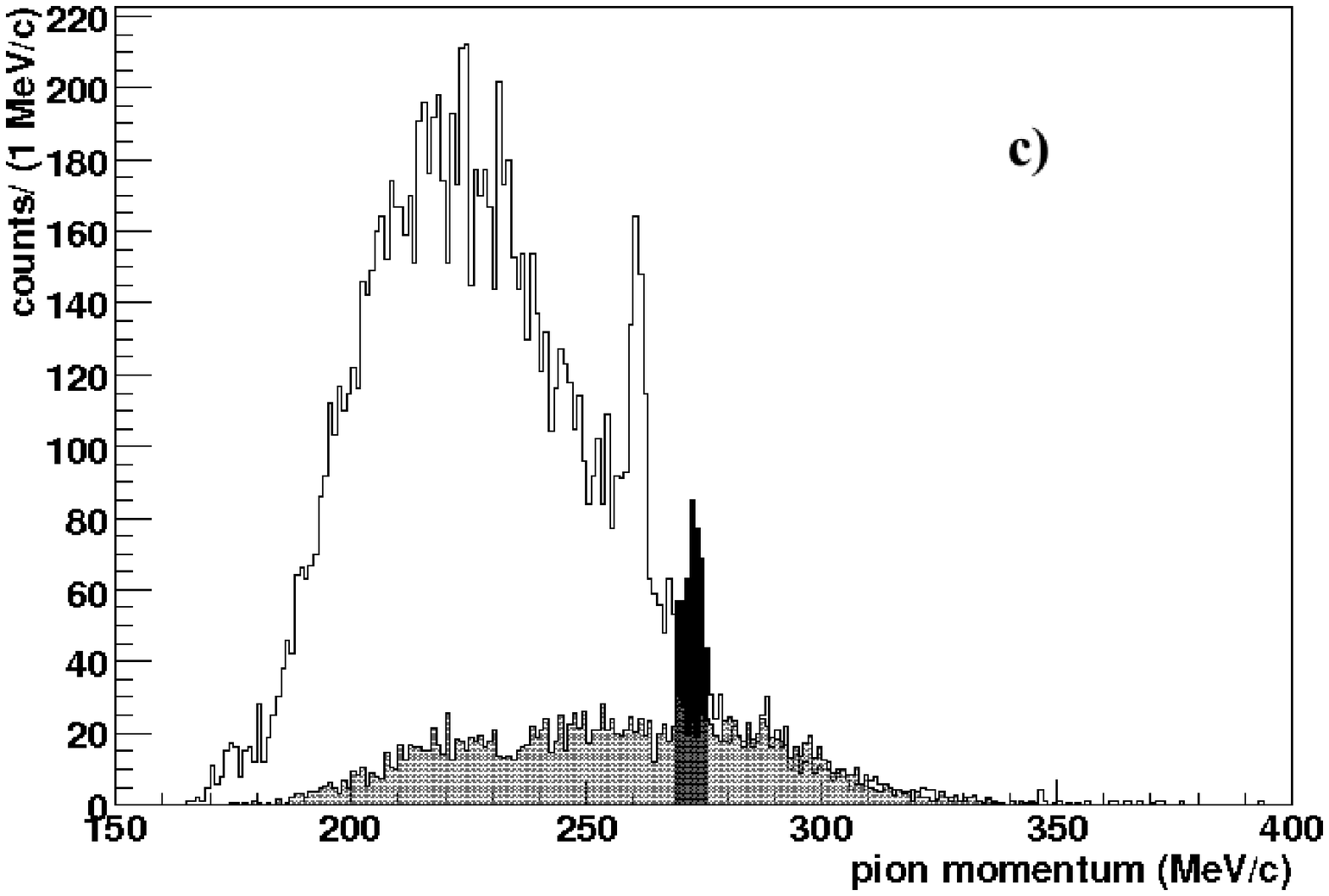}}
\caption{Momentum distribution of negative pions detected in coincidence with a proton from a)$\mathrm{^{6}Li}$ b)$\mathrm{^{7}Li}$ and c)$\mathrm{^{12}C}$ targets. The black regions corresponds respectively to  a)$\mathrm{^5_{\Lambda}He}$, b) $\mathrm{^7_{\Lambda}Li}$ and c)  $\mathrm{^{12}_{\Lambda}C}$  g.s. formation. The inset in b) shows the region corresponding to the formation of ($\mathrm{^5_{\Lambda}He}$+d) system (black area) with the corresponding calculated background (grey area).}
\label{f:coinc}
\end{center}
\end{figure}

Fig.\ref{f:coinc} shows the $\pi^{-}$ spectra when a proton, reconstructed accepting both {\it long} and {\it short tracks}, is required to be emitted in coincidence from the primary vertex of interaction. The black areas represent the $\pi^{-}$ momentum region (6 MeV/c wide) corresponding to the formation of the aforementioned  $\mathrm{^5_{\Lambda}He}$,  $\mathrm{^7_{\Lambda}Li}$ and  $\mathrm{^{12}_{\Lambda}C}$ Hypernuclei.

\begin{figure}[htb]
\begin{center} 
\begin{tabular}{cc}
\resizebox {7cm}{!}{\includegraphics{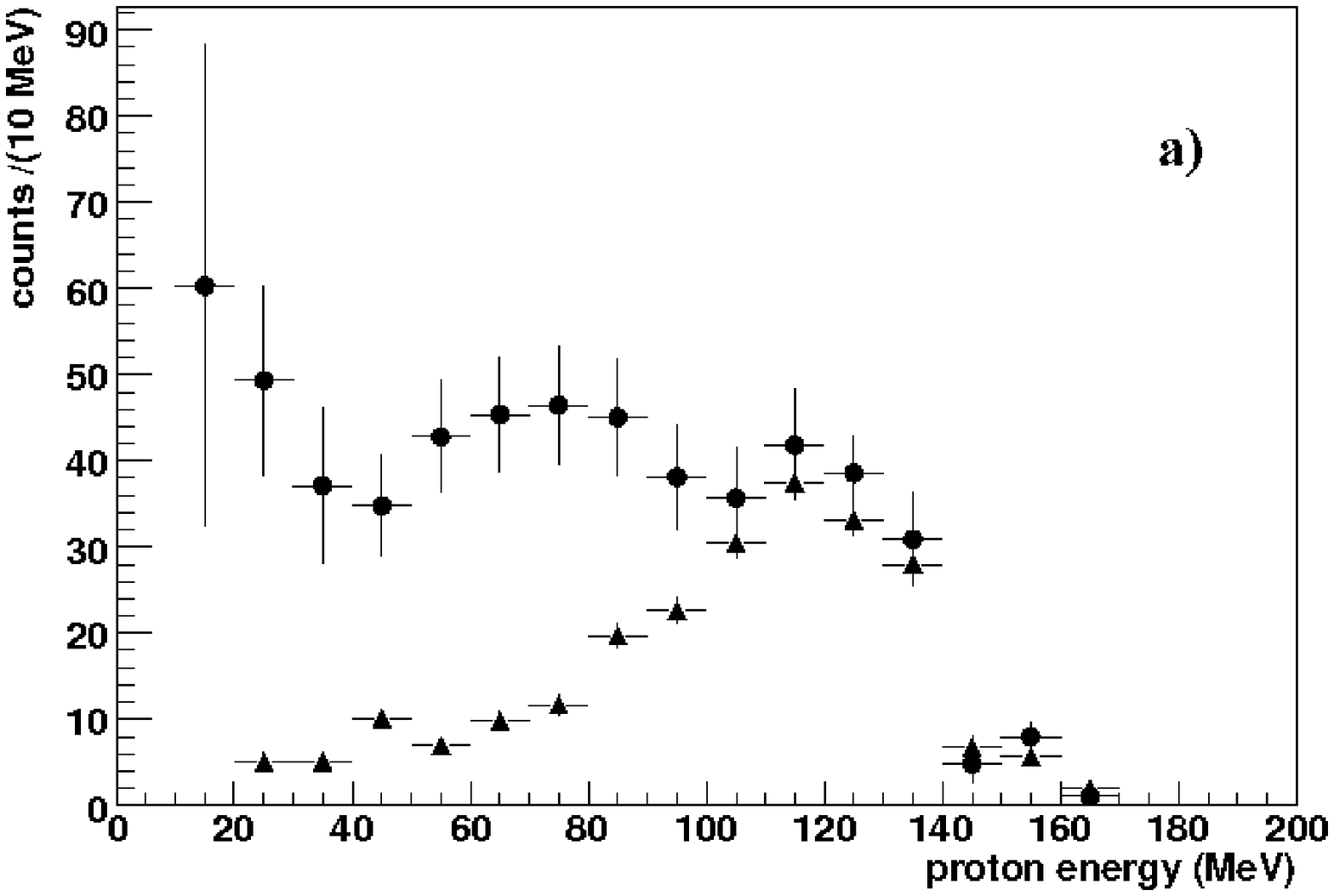}} & \resizebox {7cm}{!}{\includegraphics{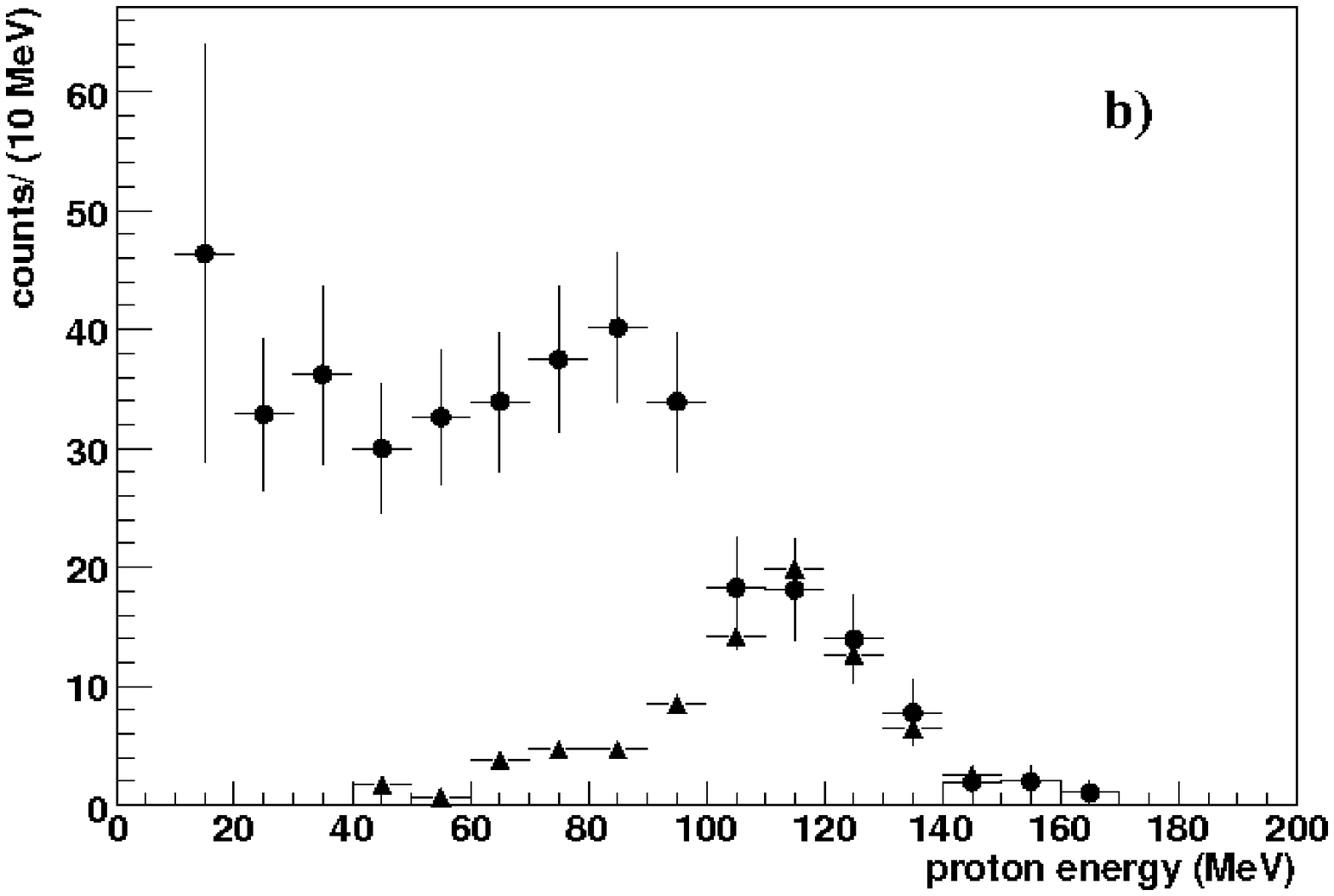}}
\end{tabular}
\resizebox {7cm}{!}{\includegraphics{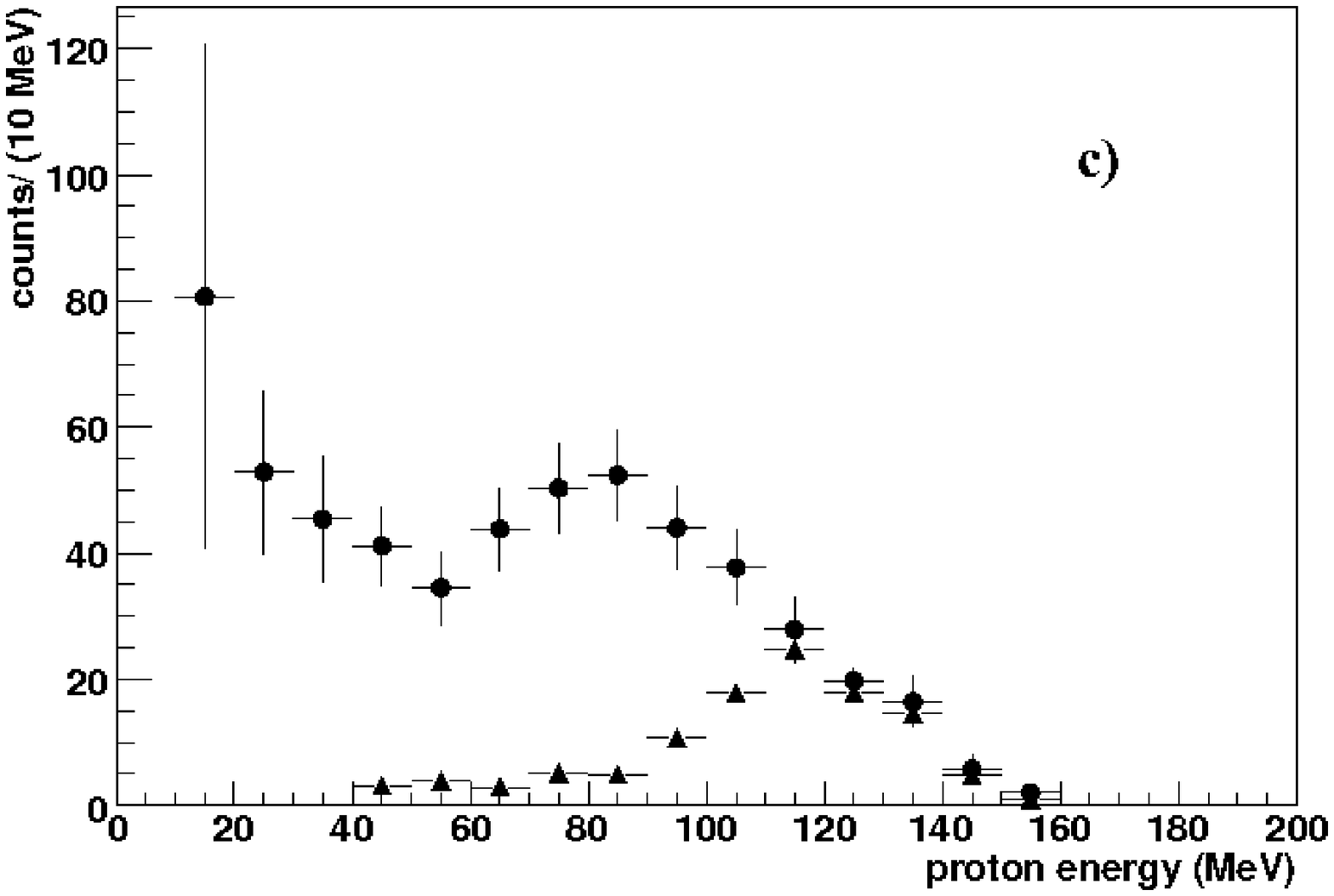}}
\caption{Experimental, acceptance corrected, proton spectra for a)$\mathrm{^5_{\Lambda}He}$, b) $\mathrm{^7_{\Lambda}Li}$ and  c) $\mathrm{^{12}_{\Lambda}C}$ (dots); triangles:energy spectrum of the protons coming from the background reaction calculated as described in the text.}
\label{f:proto}
\end{center}
\end{figure}

Fig.\ref{f:proto} shows the acceptance corrected proton energy spectra (dots) detected in coincidence with a $\pi^{-}$ of a momentum in the above mentioned windows.

\begin{figure}[htb]
\begin{center} 
\begin{tabular}{cc}
\resizebox {7cm}{!}{\includegraphics{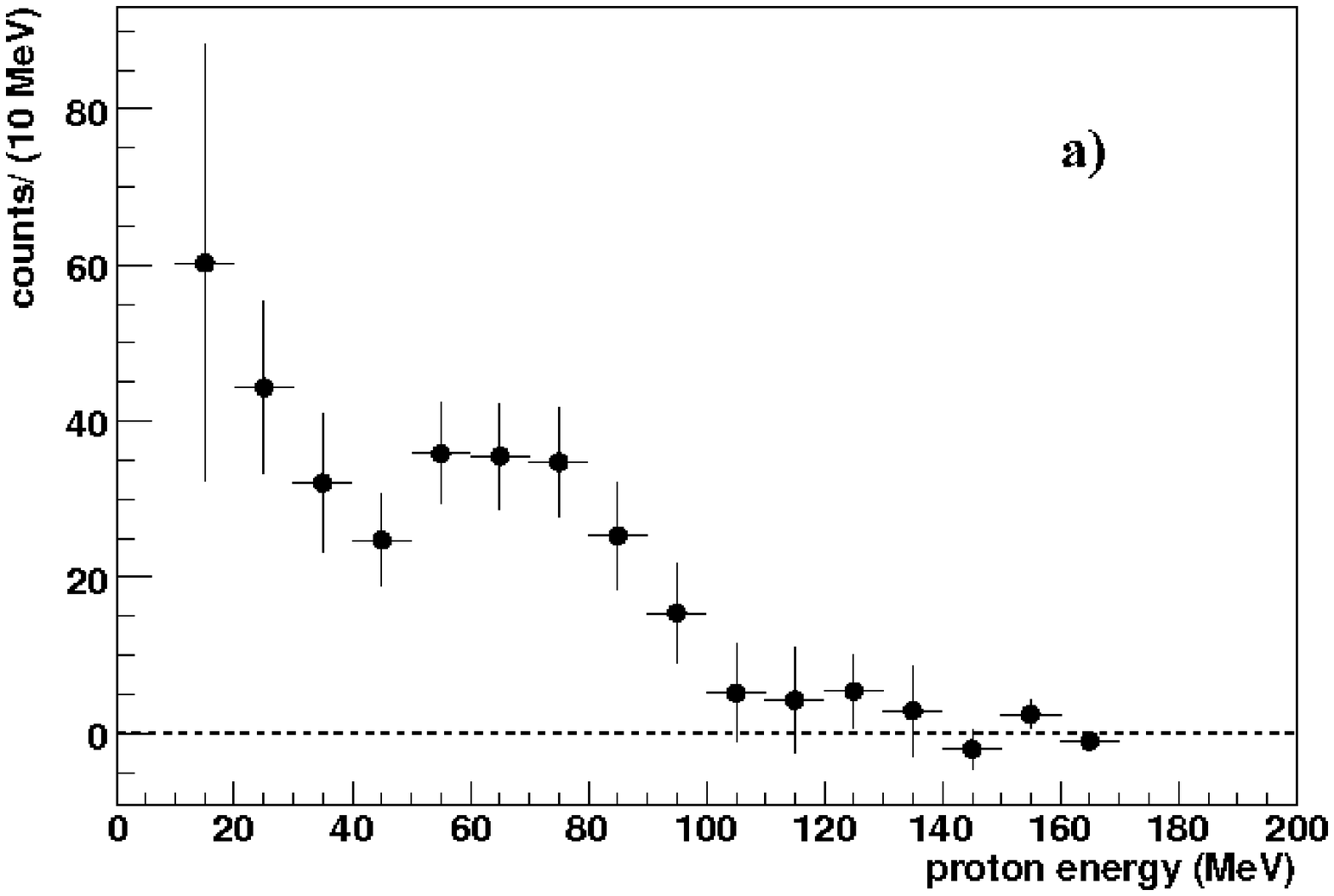}} & \resizebox {7cm}{!}{\includegraphics{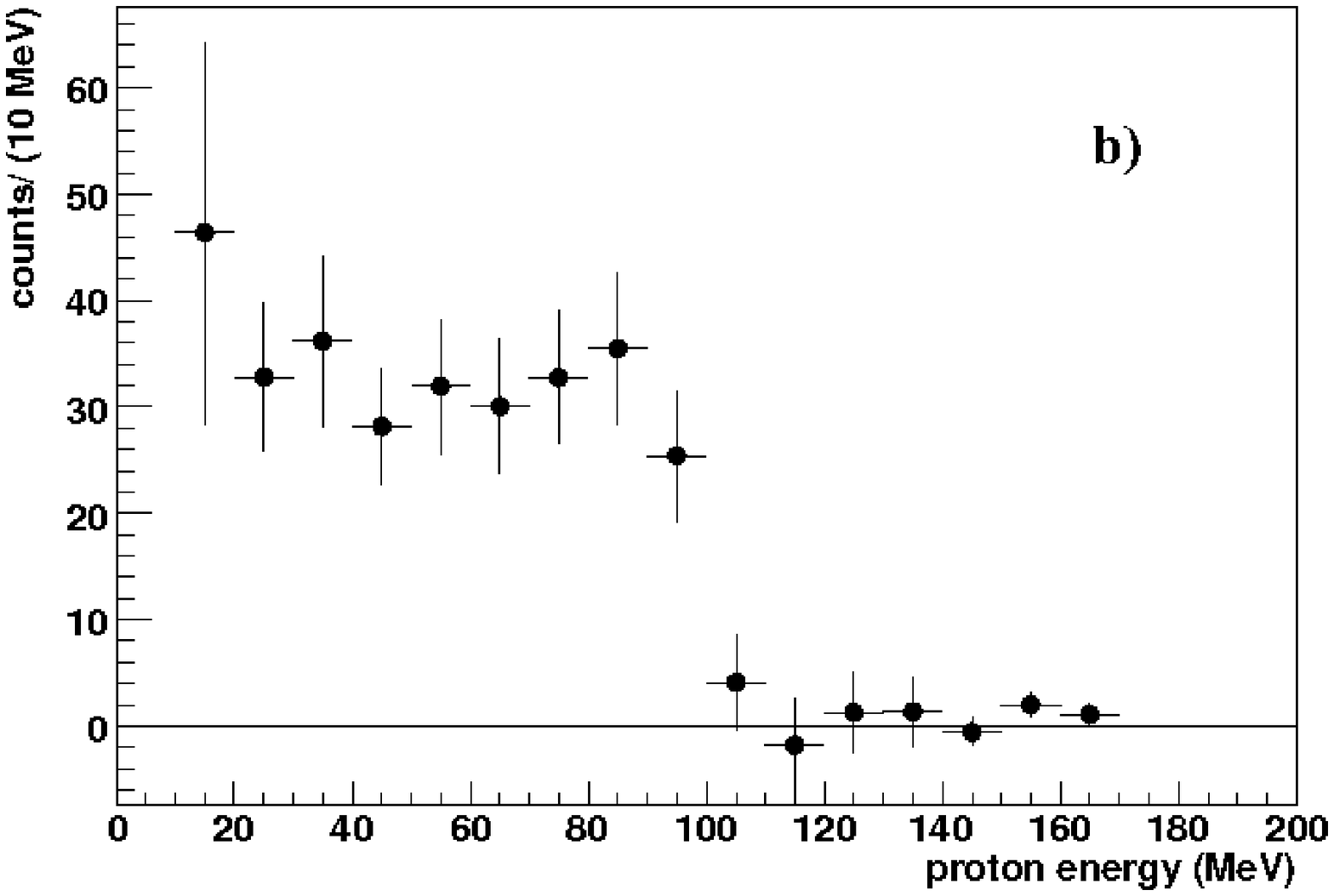}}
\end{tabular}
\resizebox {7cm}{!}{\includegraphics{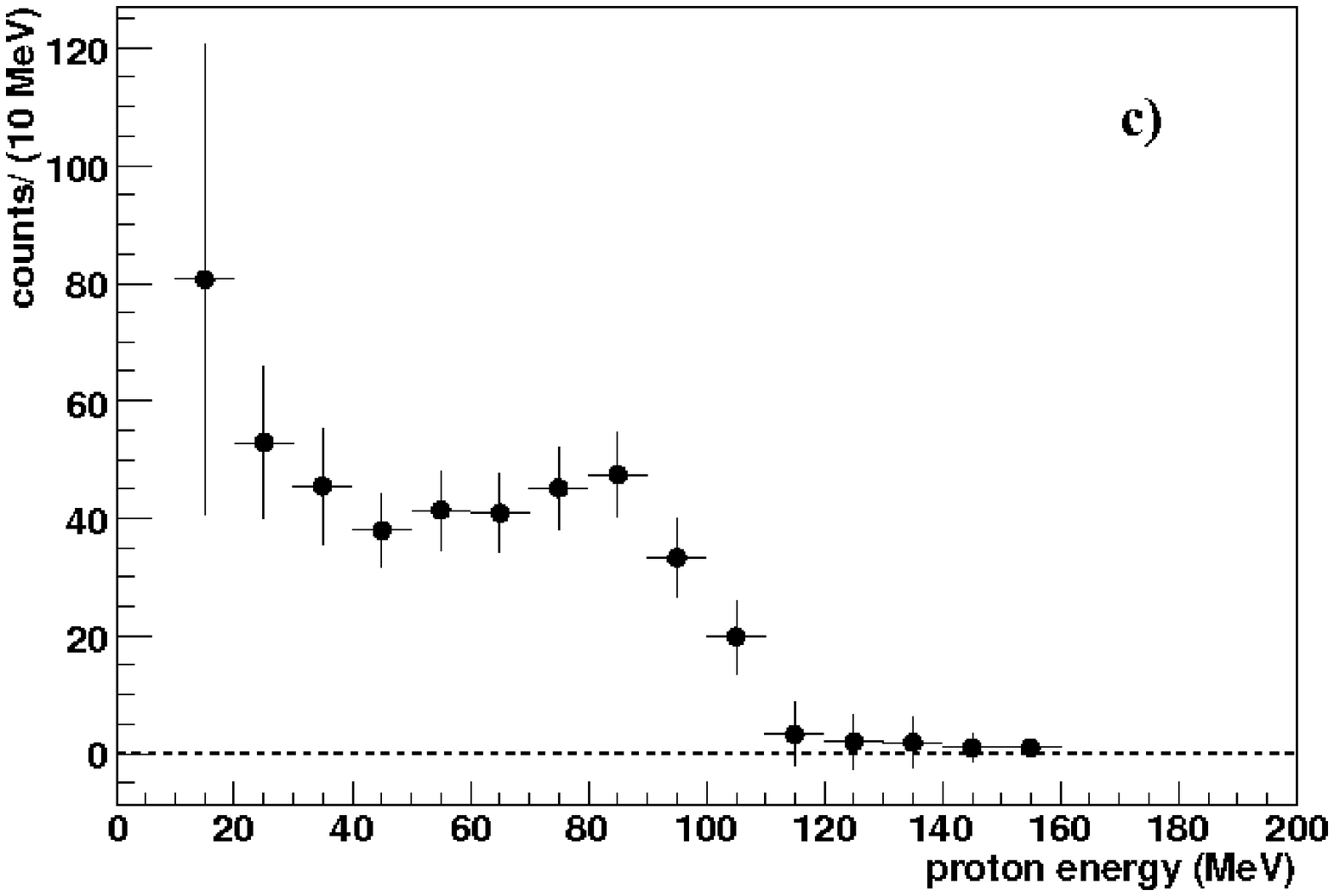}}
\caption{Proton energy spectrum after the background subtraction from NMWD of a)$^{5}_\Lambda$He, b) $^{7}_\Lambda$Li and c)$^{12}_\Lambda$C}
\label{f:segnale}
\end{center}
\end{figure}

The raw proton spectra exhibit a similar shape for $\mathrm{^7_{\Lambda}Li_{g.s.}}$  and $\mathrm{^{12}_{\Lambda}C_{g.s.}}$ (Fig.\ref{f:proto}b and \ref{f:proto}c respectively), with a maximum at about 80 MeV, roughly half of the Q--value for the $\Lambda \cal{N} \rightarrow \cal{N} \cal{N}$ reaction (156 MeV), whereas that for  $\mathrm{^5_{\Lambda}He_{g.s.}}$ shows a camel-back double humped structure, with a second maximum around 120 MeV. $\pi^{-}$'s with momenta larger than the kinematical limit for the Hypernuclei production reaction  are mainly due to the $K^{-} ({\it np}) \rightarrow \Sigma^{-} {\it p}$ absorption on correlated nucleons in the absorbing nucleus, followed by $\Sigma^{-}$ decay in flight (\ref{eq2}) occurring at distances smaller than 3 mm. Out of the several processes that lead to a continuous spectrum of $\pi^{-}$ observed in $K^{-}$ absorption at rest, this one is the only one affecting the region of the bound states of Hypernuclei. It was suitably parametrized \cite{ref:FINUDAcarbonio} and the grey areas below the spectra of Fig.\ref{f:coinc}(b,c) represent the result of a Monte Carlo simulation corresponding to  4 $\times$ $10^{6}$ events. The simulated events were selected with the same criteria of the real data.\\ 
The errors on each bin of the simulated spectrum are one order of magnitude lower than those on the measured spectrum. 
The simulated spectra were normalized to the experimental ones for $\pi^{-}$ momenta larger than 278 MeV/c ($^{7}_\Lambda$Li) and 276 MeV/c ($^{12}_\Lambda$C). The agreement between the simulated and experimental spectra is very good ($\chi^2$/d.o.f equal to 1.3 for $^{7}_\Lambda$Li and $\chi^2$/d.o.f= 1.2 for $^{12}_\Lambda$C).  \\
The triangles in Fig.\ref{f:proto} show the  acceptance corrected proton spectra coming from the two--nucleon absorption mechanism, evaluated by the Monte Carlo simulation program and corresponding to $\pi^{-}$ with momentum in the ranges (270--276) MeV/c for $^{12}_\Lambda$C and (272--278) MeV/c for $^{7}_\Lambda$Li. The final proton spectra  after the background subtraction are shown in Fig \ref{f:segnale}(b,c). 

The subtraction of the two nucleon absorption mechanism for $^{6}$Li cannot be done satisfactorily by using the model valid for the other nuclei. Indeed it is well known that in many reactions $^{6}$Li behaves like an ($\alpha$+d) molecule, and this cluster substructure has to be taken into account in the simulation.
We verified this feature in a previous work \cite{ref:Ala} and we modeled the  $K^{-}$ ({\it np}) absorption process taking into account the momentum distribution of the deuteron inside $^{6}$Li taken from \cite{ref:YamaLi}. The result of this simulation is represented by the grey area of Fig\ref{f:coinc}a, normalized to the experimental spectrum beyond 278 MeV/c. A reasonable agreement is observed ($\chi^2$/{d.o.f= 1.4). The simulated proton spectrum, acceptance corrected, corresponding to  $\pi^{-}$ emitted in the (272--278) MeV/c range (black area of Fig.\ref{f:coinc}a is represented by the triangles of Fig.\ref{f:proto}a and the final background subtracted spectrum is represented by Fig\ref{f:segnale}a.


A closer inspection of the $\pi^{-}$ momentum spectrum observed for $^{7}$Li with a proton coincidence shows a hint for a peak, centered around 269 MeV/c, i.e. 6 MeV/c below the $\pi^{-}$ peak corresponding to the 
 $^{7}_\Lambda$Li formation, and labeled by the black area in the inset of Fig.\ref{f:coinc}b. The hypernuclear systems that can be produced stopping a $K^{-}$ in a $^{7}$Li target are the following: $^{7}_\Lambda$Li, ($^{6}_\Lambda$H + p), ($^{5}_\Lambda$He + d), ($^{4}_\Lambda$He + t) and ($^{3}_\Lambda$H + $\alpha$). From the Hypernuclear Mass Tables we find that, after $^{7}_\Lambda$Li with a mass m= 6711.6 MeV the lowest hypernuclear system is found for the ($^{5}_\Lambda$He + d) system with m=6715.57 MeV. Therefore the black peak can be assigned to the reaction:

\begin{equation}
K^{-}_{stop} + \mathrm{^{7}Li} \rightarrow \mathrm{^{5}_\Lambda He} + d + \pi^{-}
\label{eq:eq13}
\end{equation} 

The maximum momentum of the $\pi^{-}$ emitted in (\ref{eq:eq13}), back--to--back to the ($^{5}_\Lambda$He + d) system is 272.67 MeV/c.
Looking at the inset of Fig.\ref{f:coinc}b the end point of this second peak is exactly at 272 MeV/c. Furthermore, by calculating the difference in binding energy  $\mathrm{B_{\Lambda}}$ between the ($^{5}_\Lambda$He + d) and $^{7}_\Lambda$Li formation corresponding to the emission of a $\pi^{-}$ with respectively 269 and 275 MeV/c momentum  (central values of the two peaks in the experimental spectrum) a value $\Delta B_{\Lambda}$=3.98 MeV was obtained, fully consistent with the experimental determination  \cite{ref:Tamu}. This observation suggested us that the proton spectrum measured in coincidence with a  $\pi^{-}$ in the momentum region of the  $\pi^{-}$ peak labeled by the black area of the inset of Fig.\ref{f:coinc}b is that due to the  NMWD of $^{5}_\Lambda$He. The experimental proton spectrum, acceptance corrected, is shown in Fig.\ref{f:pLi7}a. The dots represents the simulated proton spectra obtained as previously described. The final background subtracted spectrum is shown by Fig.\ref{f:pLi7}b.

\begin{figure}[htb]
\begin{center}
\begin{tabular}{cc}
\resizebox{7cm}{!}{\includegraphics{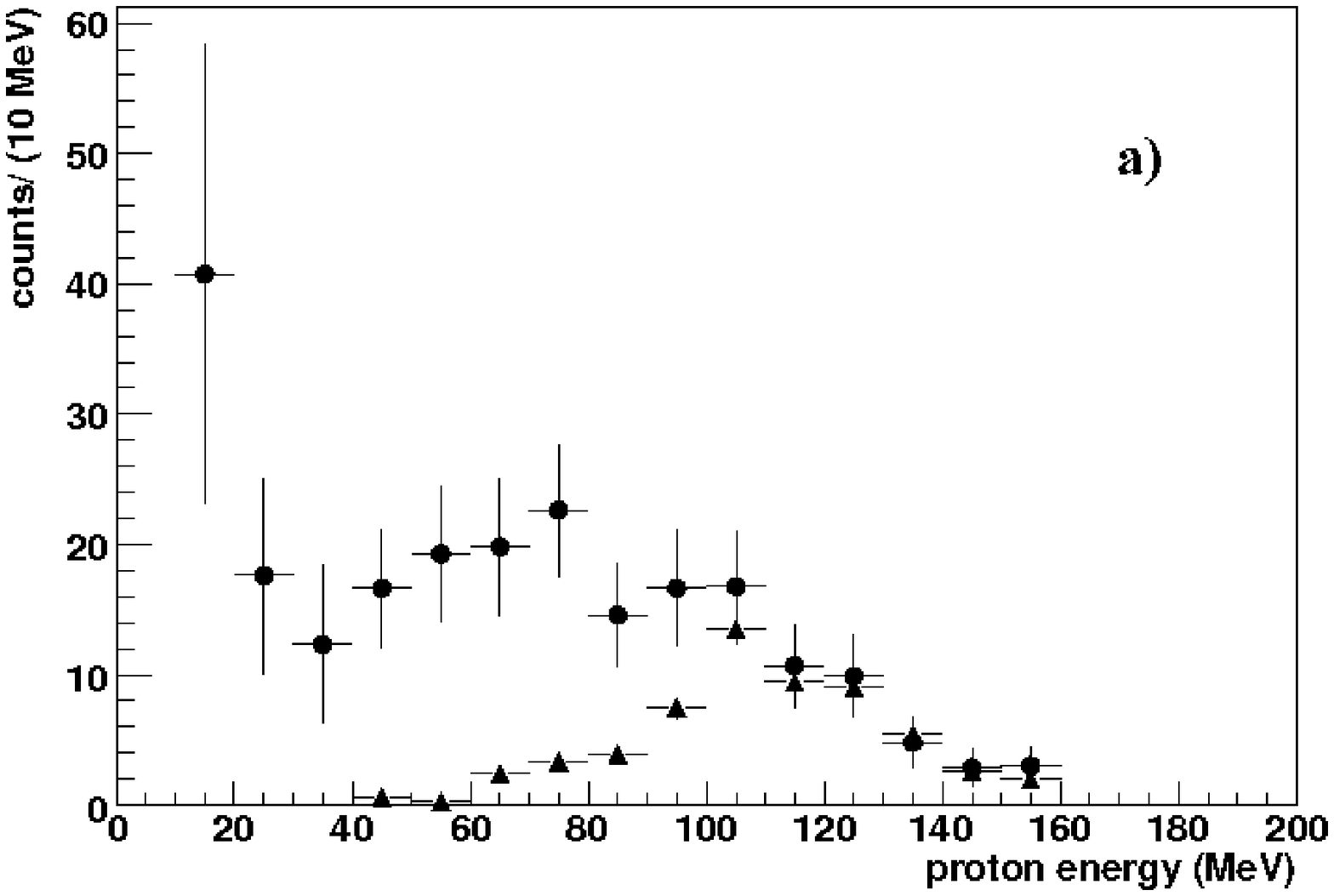}} & \resizebox{7cm}{!}{\includegraphics{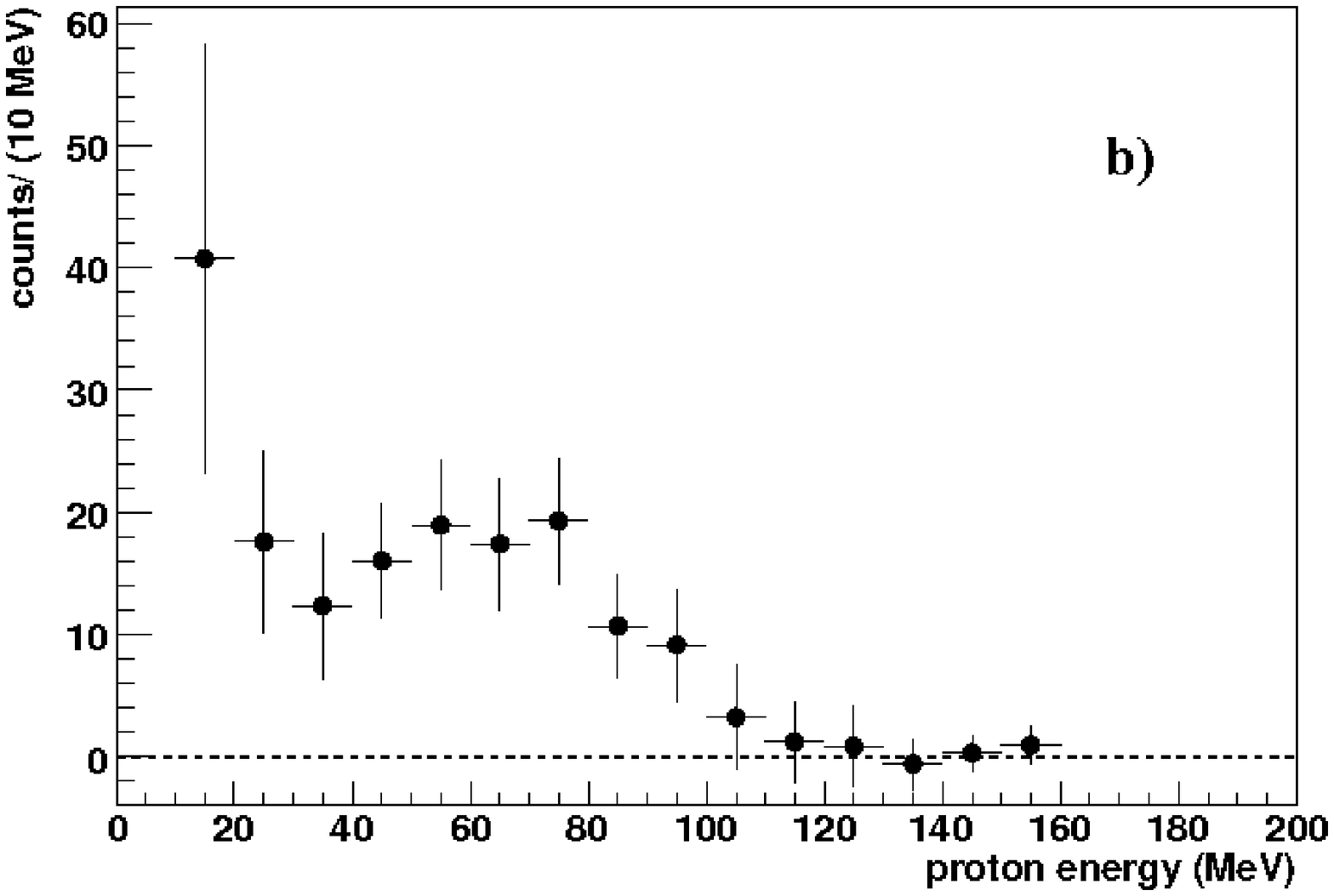}}
\end{tabular}
\caption{a): experimental, acceptance corrected, proton spectrum obtained from NMWD of $^{5}_\Lambda$He as obtained from the $^{7}$Li targets (dots); triangles: acceptance corrected  proton spectrum coming from the background reaction calculated as described in the text. b):Proton energy spectrum from NMWD of $^{5}_\Lambda$He ($^{7}$Li targets) after the background subtraction.}  
\label{f:pLi7}
\end{center}
\end{figure}

The comparison of the proton spectra corresponding to the NMWD of $^{5}_\Lambda$He out from the $^{6}$Li and $^{7}$Li targets (Fig.\ref{f:segnale}a and Fig\ref{f:pLi7}b respectively)  shows that they are the same within the errors. The Kolmogorov-Smirnov (K--S) test applied to these data gives a value of compatibility  at a confidence level (C.L) of 95\%, meaning that the spectra are fully compatible and can be added safely. The sum of the spectra is shown in Fig.\ref{f:sumeli}

\begin{figure}[htb]
\begin{center}
\resizebox {7cm}{!}{\includegraphics{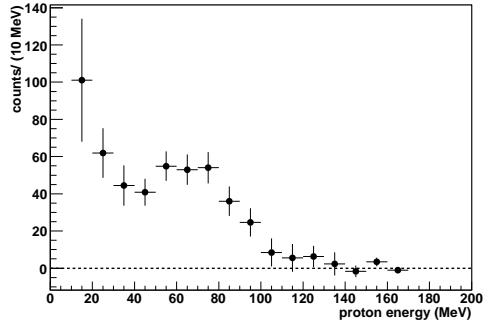}}
\caption{Total proton energy spectrum of  $^{5}_\Lambda$He obtained adding the spectrum from $^{7}$Li targets analysis (Fig.\ref {f:pLi7}b) and the one from the analysis of $^{6}$Li targets  (Fig.\ref{f:segnale}a).}\label{f:sumeli}
\end{center}
\end{figure}

\section{Discussion of results and conclusions}

As mentioned in the Introduction, proton spectra from NMWD of $^{5}_\Lambda$He and $^{12}_\Lambda$C were already studied experimentally \cite{ref:BNL91,ref:BNL95,ref:Okal,ref:sato} and theoretically \cite{ref:parreno,ref:alberico,ref:Bau,ref:Garba}. 
For $^{7}_\Lambda$Li there are no measurements nor theoretical calculations, in spite of the fact it is the best known Hypernucleus  from the spectroscopic point of view \cite{ref:Tamu}. The proton spectra from NMWD of $^{5}_\Lambda$He, fig.\ref{f:sumeli}, $^{7}_\Lambda$Li, fig.\ref{f:segnale}b, and $^{12}_\Lambda$C, fig.\ref{f:segnale}c, show 
a similar shape, i.e. a peak around 80 MeV, corresponding to about a half the Q-value for the free $\Lambda p \rightarrow n p$ weak reaction, with a low energy rise, 
due to the final state interactions (FSI) and/or to two nucleon induced weak decays \cite{ref:alberico,ref:Bau,ref:Garba}. 

This observation doesn't allow us to extract the proton decay width $\Gamma_p$ from the experimental data. We may give the ratio $R_p$ of the total number of protons with 
energy larger than  15 MeV to the total number of produced Hypernuclei. We evaluated $R_p$ by the simple formula:
\begin{equation}
R_p = \frac{N^{detected}_{p}}{N^{detected}_{hyp} \epsilon _p}
\label{eq:r_p}
\end{equation}  
in which $N^{detected}_{p}$ and $N^{detected}_{hyp}$  are, respectively, the number of protons and of hypernuclei derived from the experimental spectra after the 
background subtraction and $\epsilon _p$ is the global acceptance of the FINUDA apparatus for protons. It was evaluated as outlined in Section 2. We underline 
the fact that we could safely use the very simple relationship (\ref{eq:r_p}) for the evaluation of $R_p$ thanks to the use of very thin targets and of a 
transparent detector. On the contrary, in previous experiments \cite{ref:BNL91,ref:BNL95,ref:Okal,ref:sato} thick targets were used to improve the statistics with the 
consequence that many corrections  had to be applied. The results are reported in Table \ref{f:tab7}. The errors reported in Table\ref{f:tab7} are the statistical ones. 
We estimated that the systematical error is lower than $5\%$.
\begin{table}[htdp]
\begin{center}
\begin{tabular}{|c|c|c|c|}
\hline 
\textbf{Target } & \textbf{Hypernucleus}& \textbf{$R_{p}$} \\
                                        & &(energy$\ge$15 MeV) \\
\hline  
\hline 
$^{12}$C & $^{12}_\Lambda$C & 0.43$\pm$0.07 \\
$^{6}$Li & $^{5}_\Lambda$He & 0.28$\pm$0.09 \\
$^{7}$Li & $^{7}_\Lambda$Li & 0.37$\pm$0.09 \\
$^{7}$Li & $^{5}_\Lambda$He & 0.21$\pm$0.12 \\
weighted mean & & \\ 
of $^{6}$Li  and $^{7}$Li   & $^{5}_\Lambda$He & 0.25$\pm$0.07 \\
\hline 
\end{tabular}
\caption{$R_{p}$ values calculated by means of the relation (\ref{eq:r_p}). The reported errors are statistical only.} 
\label{f:tab7}
\end{center}
\end {table}

Fig.\ref{f:He}a shows the comparison of our spectrum for $^{5}_{\Lambda}$He with the one by Okada et al.\cite{ref:Okal}. The two spectra were normalized 
beyond 35 MeV (the proton energy threshold of \cite{ref:Okal}). The (K--S) test applied to the two data sets provides a C.L. of $75\%$. 
\begin{figure}[htb]
\begin{center}
\begin{tabular}{cc}
\resizebox {7cm}{!}{\includegraphics{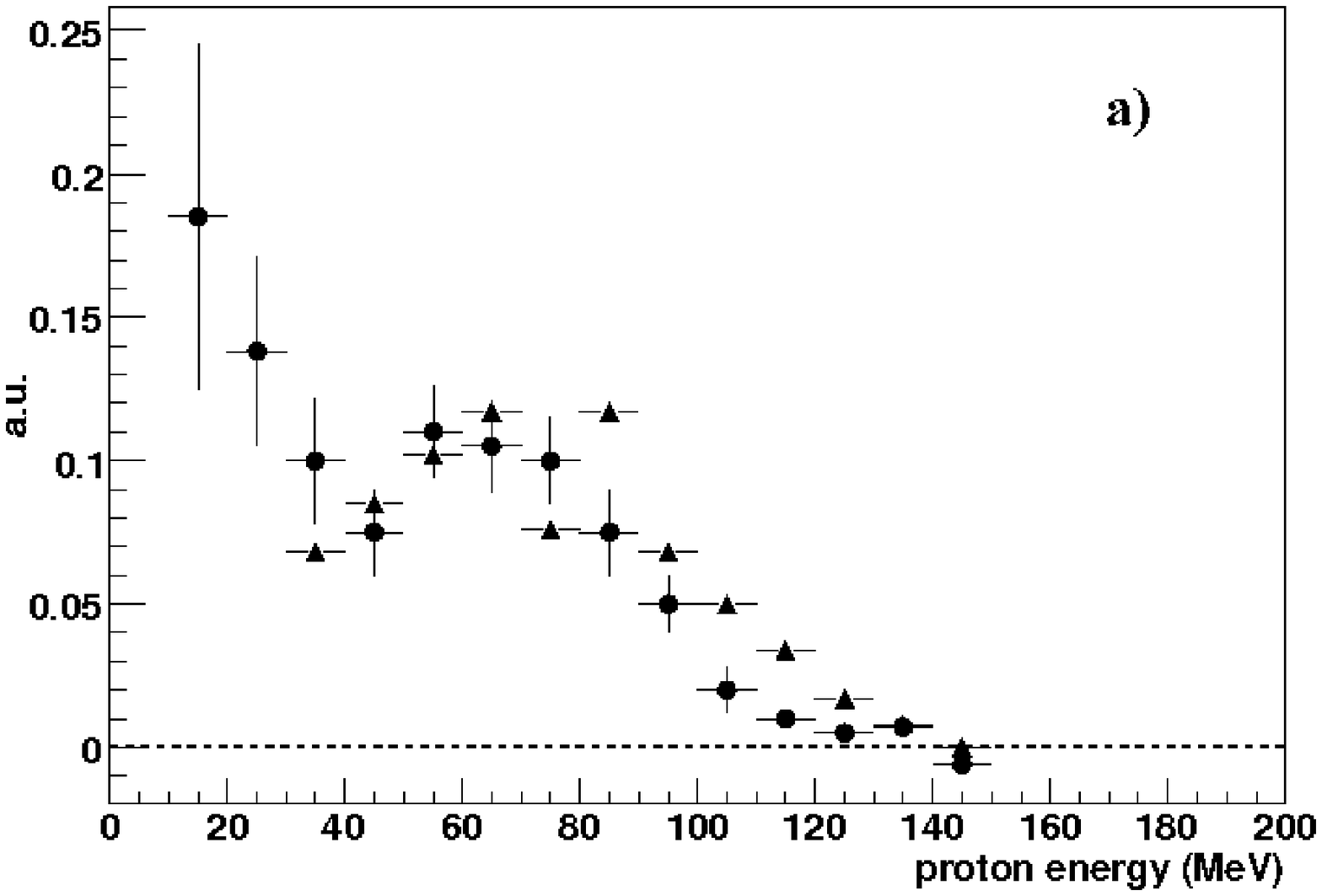}} \resizebox {7cm}{!}{\includegraphics{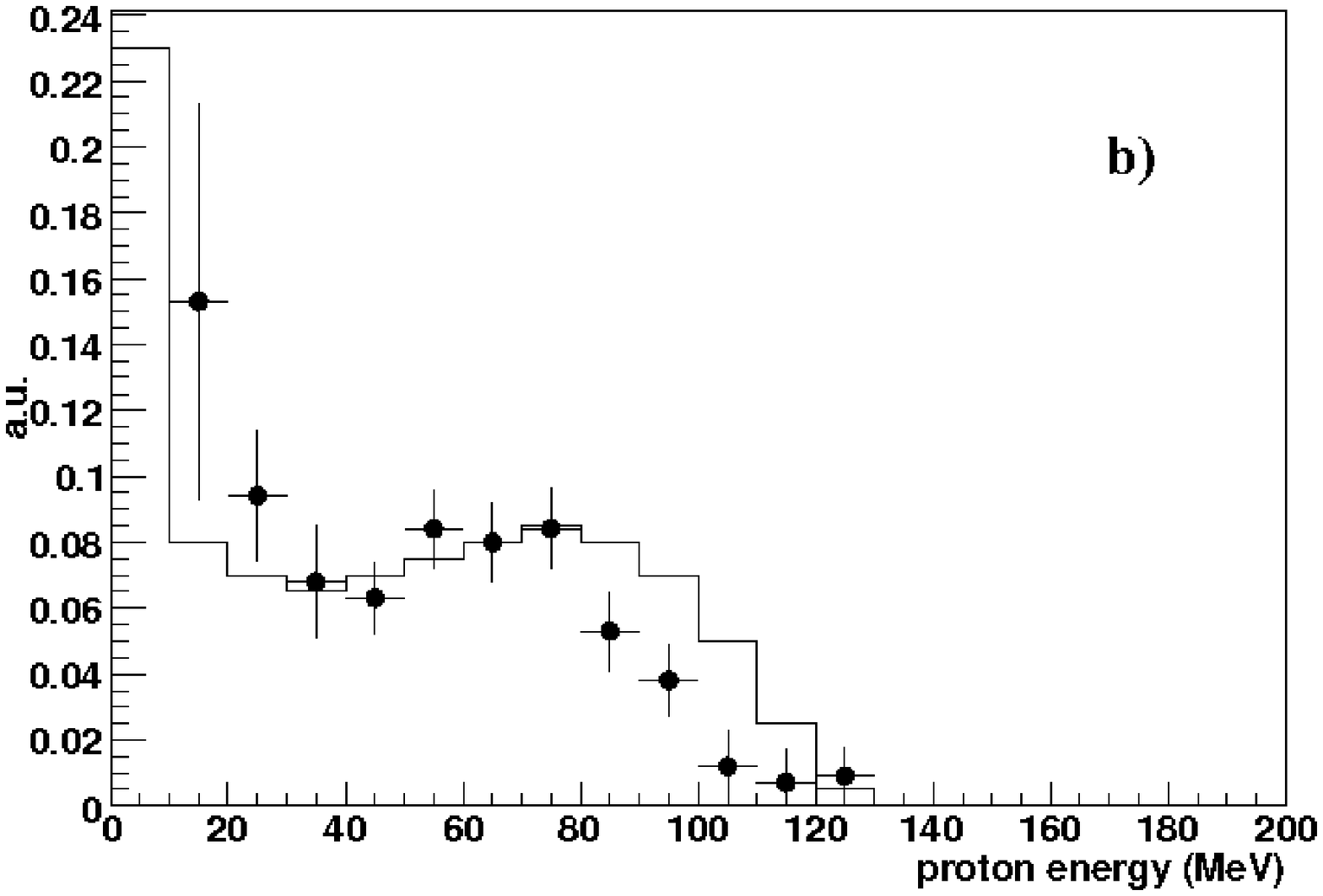}} 
\end{tabular}
\caption{a)dots: FINUDA proton spectrum from proton--induced NMWD for $^{5}_{\Lambda}$He; triangles: result achieved for the $^{5}_{\Lambda}$He at the KEK experiments; the two spectra are normalized to areas beyond 35 MeV. b) Dots: experimental proton spectra of $^{5}_\Lambda$He; continous histogram: theoretical calculation for the proton energy spectrum of $^{5}_\Lambda$He performed with the addition of the FSI contribution. The two spectra are normalized  beyond 15 MeV.}
\label{f:He}
\end{center}
\end{figure}

Fig.\ref{f:He}b shows the comparison of our spectrum with the theoretical one calculated by Garbarino et al.\cite{ref:Garba}. The two spectra were normalized 
beyond 15 MeV (our proton energy threshold). The K--S test applied to the two data sets provides a C.L. of $80\%$. We may conclude that there is a disagreement 
between the two experiments and with the theory, even though not so severe.

The situation for $^{12}_{\Lambda}$C is completely different.  Fig \ref{f:carbo}a shows the comparison of our spectrum with that by Okada et al.\cite{ref:Okal}: the two 
spectra were normalized  beyond 35 MeV.
\begin{figure}[htb]
\begin{center} 
\begin{tabular}{cc}
\resizebox {7cm}{!}{\includegraphics{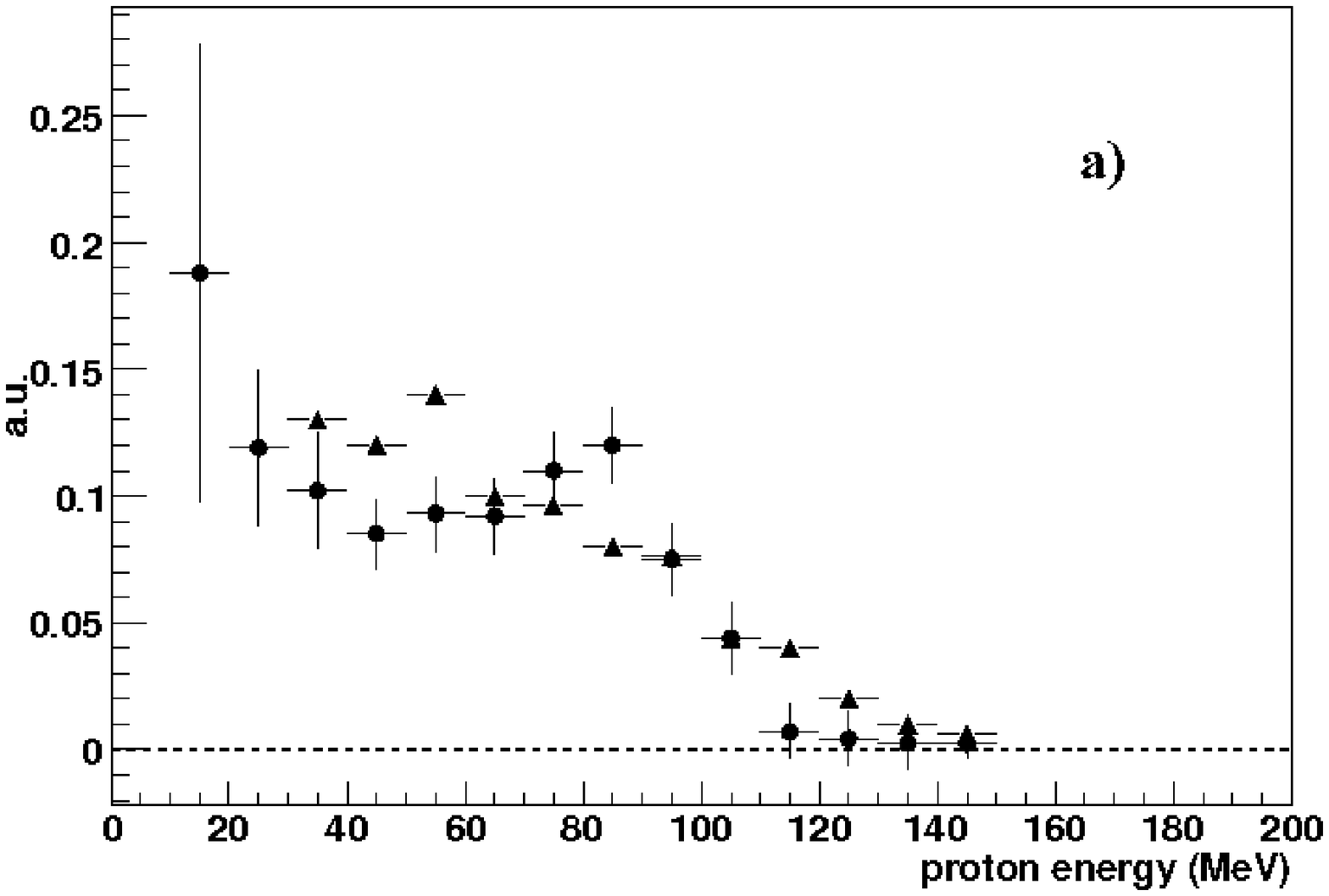}} & \resizebox {7cm}{!}{\includegraphics{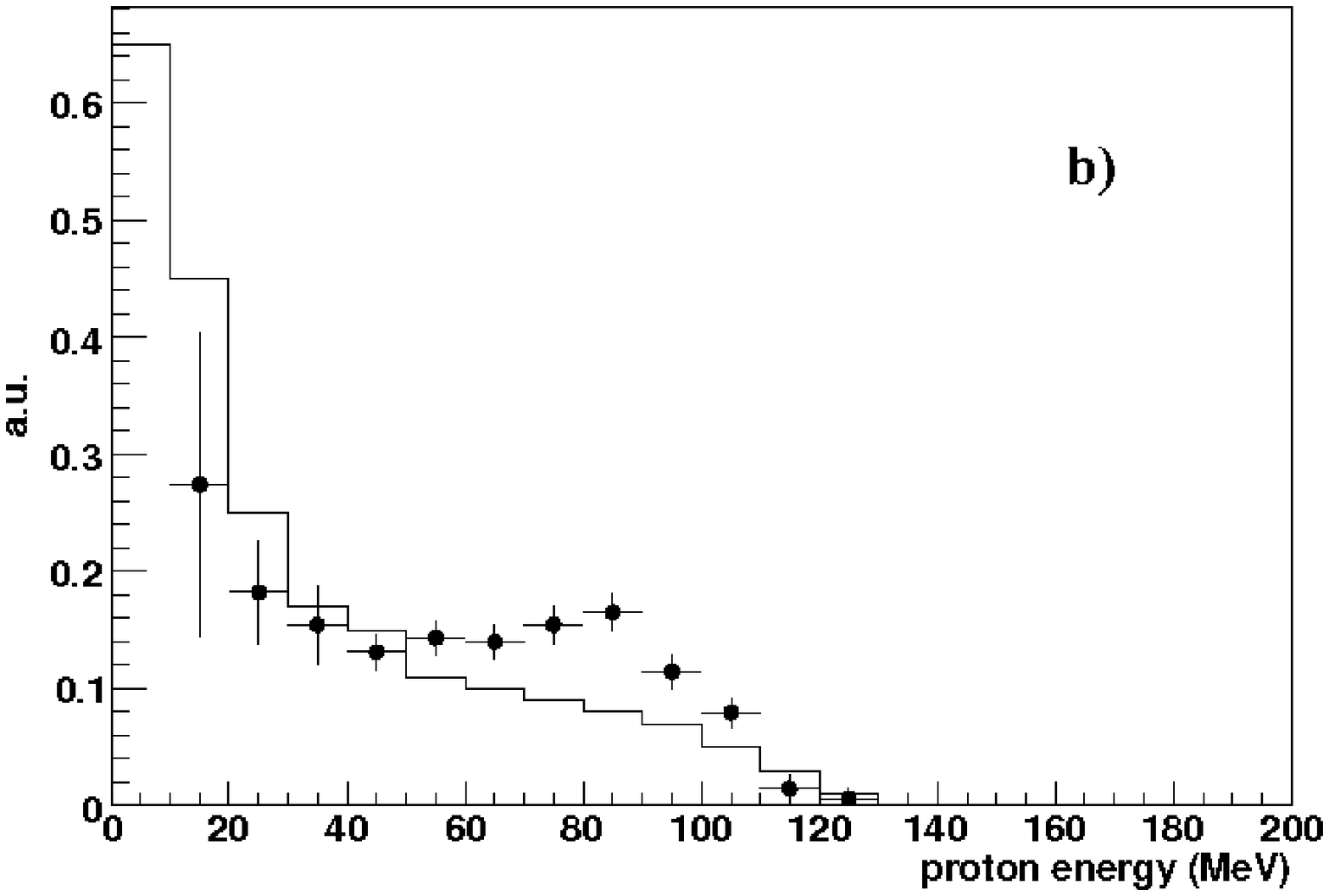}}
\end{tabular}
\caption{ Comparison between the proton spectra from NMWD of $^{12}_{\Lambda}$C measured by a) FINUDA and Okada et al.\cite{ref:Okal} normalized beyond 35 MeV; b) comparison between the FINUDA spectrum and the theoretical calculation performed by Garbarino et al. \cite{ref:Garba} normalized beyond 15 MeV.}
\label{f:carbo}
\end{center}
\end{figure}

The K--S test applied  to the two data sets provides a C.L. of $20\%$.  Fig \ref{f:carbo}b shows the comparison of our spectrum with the theoretical one 
calculated by Garbarino et al.\cite{ref:Garba}. The two spectra were normalized beyond 15 MeV. The K--S test applied to the two data sets provides a C.L. of $5\%$. The conclusion is that there is a strong disagreement between the two experiments and with the theory. 

Concerning the discrepancy between the two sets of experimental data, we may remark that in \cite{ref:Okal} the proton energy was measured by a combination of 
time-of-flight and total energy deposit measurements. The energy loss inside the thick targets was corrected event-by-event. The energy resolution became poorer in the 
high energy region, especially above 100 MeV, with the consequence that the spectra could be strongly distorted. On the contrary with FINUDA the proton momenta are 
measured by means of a magnetic analysis, with an excellent resolution ($2\%$ FWHM) and no distortion at all on the spectra is expected, in particular in the high energy 
region. We believe that our spectrum is the genuinely undistorted one, even though with a limited statistics.

The spectra for $^{5}_{\Lambda}$He, $^{7}_{\Lambda}$Li and $^{12}_{\Lambda}$C look quite similar, in spite of the large mass number difference of these nuclei. 
If the low energy rises were predominantly due to FSI effects, one should naturally  expect that the broad peak structure at 80 MeV (coming from clean $\Lambda p 
\rightarrow n p$ weak processes broadened by the Fermi motion of nucleons) would be smeared out for the heavier nuclei.

Another effect that should be the origin of the low energy rise is a large contribution of the two-nucleon induced weak process $\Lambda n p \rightarrow n n p$
\cite{ref:alberico,ref:Garba}. If the weak decay Q-value (160 MeV) is shared by three nucleons, a low energy rise may exist even for the very light s-shell hypernuclei. 
Our data seem to agree with the hypothesis of a substantial contribution of the two--nucleon induced NMWD. Indeed recently  Bhang et al.\cite{ref:bhan} suggested a contribution of the two-nucleon induced weak decay process as large as $40\%$ of the total NMWD width for $^{12}_{\Lambda}$C.
However, in order to 
strengthen this conclusion, there is an urgent need for more data on more medium-A nuclei with much more statistics. 

We are grateful to Dr. G. Garbarino for useful discussions and a critical reading of the manuscript.

\end{document}